\documentclass[useAMS,usenatbib,usegraphicx]{mn2e}
\usepackage{bm}
\usepackage{times}

\title[Dust growth through accretion and coagulation]
{Dust growth in the interstellar medium: How do
accretion and coagulation interplay?}
\author[Hirashita]{Hiroyuki Hirashita$^1$\thanks{E-mail:
    hirashita@asiaa.sinica.edu.tw}
\\
$^1$Institute of Astronomy and Astrophysics, Academia Sinica,
P.O. Box 23-141, Taipei 10617, Taiwan
}
\date{2012 February 6}

\pagerange{\pageref{firstpage}--\pageref{lastpage}} \pubyear{2012}

\begin{document}
\label{firstpage}
\maketitle

\begin{abstract}
Dust grains grow in interstellar clouds by accretion
and coagulation. In this paper, we focus on these two
grain growth processes and numerically investigate
how they interplay to increase the grain radii.
We show that
accretion efficiently depletes grains with radii
$a\la 0.001~\micron$
on a time-scale of $\la 10$ Myr in solar-metallicity
molecular clouds. Coagulation also occurs on
a similar time-scale, but accretion is more
efficient in producing a large bump
in the grain size distribution.
Coagulation further pushes the grains to
larger sizes
after a major part of the gas phase metals
are used up. Similar grain sizes are achieved
by coagulation
regardless of whether accretion takes place or
not; in this sense, accretion and coagulation
modify the grain size distribution independently.
The increase of the total dust mass in a cloud is
also investigated. We show that coagulation slightly
`suppresses' dust mass growth by accretion
but that this
effect is slight enough to be neglected in
considering the grain mass budget in galaxies.
Finally we examine
how accretion and coagulation affect the extinction
curve: The ultraviolet slope and the carbon bump
are \textit{enhanced} by accretion, while they are
flattened by coagulation.
\end{abstract}


\section{Introduction}

Dust enrichment in galaxies is one of the most
important topics in galaxy evolution. Dust grains
actually modify the spectral energy distribution
of galaxies by reprocessing stellar radiation into
far-infrared wavelengths
{\citep[e.g.][]{desert90}}. Grain surfaces also
regulate interstellar chemistry; especially,
formation of molecular hydrogen predominantly
occurs on dust grains if the interstellar medium
(ISM) is enriched with dust
{\citep[e.g.][]{cazaux04}}. These effects of dust
grains point to the importance
of clarifying the dust enrichment in galaxies
(see \citealt{yamasawa11} for a recent modeling).

Dust enrichment is governed by various processes
depending on age, metallicity, etc.\
\citep{dwek98}. Dust grains are supplied by
stellar sources such as supernovae (SNe) and
asymptotic giant branch (AGB) stars
\citep[e.g.][]{kozasa09,valiante09}.
{Dust is also destroyed by SN shocks in the ISM.}
In the Milky Way, the time-scale of dust destruction
by SN shocks is
$\mbox{a few}\times 10^8$~yr (\citealt*{jones96};
but see \citealt{jones11}), while that
of dust supply from stellar sources is longer than
1~Gyr \citep{mckee89}.
Therefore, to explain a significant amount of dust
in the ISM, it has been argued that dust grains grow
in the ISM by the accretion of metals onto grains
\citep{inoue03,draine09,zhukovska08,pipino11,valiante11,
asano12}.
Dust grains grow efficiently in molecular clouds,
where the typical number density of
hydrogen molecules is $\sim 10^3$ cm$^{-3}$
\citep{hirashita00}. Larger depletion of metal elements
in cold clouds than in warm medium
\citep{savage96} may indicate grain growth in clouds.

In dense environments, dust grains grow not only by
accretion but also by coagulation. Indeed, a deficit
of very small grains contributing to the 60 $\micron$
emission is observed around a typical density in
molecular clouds $\ga 10^3$ cm$^{-3}$, and is
interpreted to be a consequence of coagulation
\citep{stepnik03}. Thus, we should treat accretion
and coagulation at the same time, since these two
processes could compete or collaborate with
each other. Moreover, accretion
and coagulation may have different impacts on
the extinction \citep*{cardelli89}: Coagulation
shifts the grain sizes to larger ranges and
flattens the extinction curve, while
accretion also increases the extinction itself.
Which of these two effects dominates can also
be clarified by
solving accretion and coagulation simultaneously.
Thus, we examine in this paper
how these two major processes of grain growth
-- accretion and coagulation -- interplay. We also
investigate the impact of grain growth
in clouds on the extinction curve.

When the dense clouds are dispersed after their
lifetimes, the dust grains grown through
accretion and coagulation are injected into the
diffuse medium, where dust destruction by SN
shocks occurs \citep{mckee89}.
\citet{hirashita09} show that grain shattering
is also efficient in the diffuse ISM. Thus, our
calculation results in this paper, which focuses
on grain growth in clouds, can be used as
`inputs' for subsequent
grain destruction and shattering. 
SN shocks do not penetrate efficiently into
dense environments and
dust destruction by SN shocks can be neglected
in dense clouds \citep{mckee89}.
Other possible dust processing
mechanisms in dense clouds such as destruction
by protostellar jets and photo-destruction by
radiation from stars are also neglected,
since the efficiencies of these processes are not
clear yet.
Thus, we do not treat these destructive processes
and concentrate on the grain growth mechanisms
(i.e.\ accretion and coagulation)
that work in the dense ISM.

This paper is organized as follows. We explain
the formulation in Section \ref{sec:formula}, and
describe some basic results on the evolution of
grain size distribution through grain growth
(accretion and
coagulation) in individual clouds in
Section \ref{sec:result}. Based on the calculation
results, we discuss effects on the extinction
curve and implication for the galaxy evolution
in Section \ref{sec:discussion}.
Finally, Section \ref{sec:conclusion} gives the
conclusion.

\section{Formulation}\label{sec:formula}

In this section, we formulate the evolution of
grain size distribution by the accretion of metals
and the growth by coagulation in an interstellar
cloud.\footnote{Although we mainly consider a
molecular cloud for a `cloud', the formulation
is not specific to molecular clouds but applicable
to any clouds including cold neutral clouds.
Therefore, we simply call the place hosting
grain growth `cloud'.} These two processes are
simply called accretion and coagulation in
this paper. We particularly focus on the interplay
between accretion and coagulation.

Throughout this paper, we call the elements
composing grains `metals'. We only treat grains
refractory enough to survive after the dispersal of
the cloud, and do not consider volatile grains
such as water ice.
We also assume that the grains are spherical with
a constant material density $s$, so that the grain
mass $m$ and the grain radius $a$ are related as
\begin{eqnarray}
m=\frac{4}{3}\pi a^3s.\label{eq:mass}
\end{eqnarray}

We define the grain size distribution such that
$n(a,\, t)\,\mathrm{d}a$ is the number density of
grains whose radii are between $a$ and
$a+\mathrm{d}a$ at time $t$. For simplicity, we
assume that the gas density is constant and that
the evolution of grain size distribution occurs
only through accretion and coagulation.
The dust mass density, $\rho_\mathrm{d}(t)$ is
estimated as
\begin{eqnarray}
\rho_\mathrm{d}(t)=
\int_0^\infty\frac{4}{3}\pi a^3s\, n(a,\, t)
\,\mathrm{d}a.\label{eq:rho_d}
\end{eqnarray}
We adopt silicate and graphite as dominant
grain species
\citep[e.g.][]{draine84}, and treat these
two species separately to
avoid the complexity arising from compound
species.

Accretion and coagulation are separately described
in the following subsections, but we solve these
two processes simultaneously in the calculation.
{The details of the numerical schemes are
explained in Appendix \ref{app:discrete}.}

\subsection{Accretion}


Because the knowledge about chemical properties
of accretion is still poor \citep{jones11}, we simplify
the picture by assuming that grain growth is
regulated by the sticking of the key species
denoted as X (X is Si and C for silicate and graphite,
respectively; \citealt[][hereafter HK11]{hirashita11}).
The mass fraction of the key species in dust is
denoted as $f_\mathrm{X}$: $f_\mathrm{X}=0.166$
for silicate (i.e.\ a fraction of 0.166 of silicate
is composed of Si), while $f_\mathrm{X}=1$
for graphite (i.e.\ graphite is composed of only C)
(Table \ref{tab:material}).
We follow the formulation in our previous paper
(HK11; see also \citealt{evans94}),
but modified for the purpose of numerical
calculations.

Since the grain number is
conserved in accretion, the
following continuity equation in terms of $n(a,\, t)$
holds:
\begin{eqnarray}
\frac{\partial n(a,\, t)}{\partial t}+
\frac{\partial}{\partial a}
\left[ n(a,\, t)\,\dot{a}\right] =0,
\label{eq:continuity}
\end{eqnarray}
where $\dot{a}\equiv\mathrm{d}a/\mathrm{d}t$
is the growth rate of the grain radius, which
is given {by the following form (HK11):}
\begin{eqnarray}
\dot{a}=\xi (t)\, a/\tau (a).\label{eq:dadt}
\end{eqnarray}
Here the growth time-scale as a function of
grain radius, $\tau (a)$, is estimated as
\begin{eqnarray}
\tau (a)\equiv
\frac{a}{{\displaystyle
\frac{n_\mathrm{X,tot}\, m_\mathrm{X}S}
{f_\mathrm{X}s}}
\left({\displaystyle
\frac{kT_\mathrm{gas}}{2\pi m_\mathrm{X}}}
\right)^{1/2}},\label{eq:tau}
\end{eqnarray}
where $n_\mathrm{X,tot}$ is the number density
of element X in both gas and dust phases,
$m_\mathrm{X}$ is the atomic mass of X,
$S$ is the sticking probability {for
accretion}, $f_\mathrm{X}$
is the mass fraction of the key species in the
dust {(see above)}, $k$ is the Boltzmann
constant, and
$T_\mathrm{gas}$ is the gas temperature.
In equation (\ref{eq:dadt}), we also introduce the
fraction of element $X$ in gas phase:
\begin{eqnarray}
\xi (t)\equiv n_\mathrm{X}(t)/n_\mathrm{X,tot},
\label{eq:xi}
\end{eqnarray}
where $n_\mathrm{X}(t)$ is the number density of
element X in gas phase as a function of time.
Note that $\dot{a}$ is
independent of $a$. The gas-phase metals
decrease by accretion as
\begin{eqnarray}
\frac{\mathrm{d}n_\mathrm{X}(t)}{\mathrm{d}t}=-
\int_0^\infty 4\pi a^2n_\mathrm{X}(t)\left(
\frac{kT_\mathrm{gas}}{2\pi m_\mathrm{X}}
\right)^{1/2}S\, n(a,\, t)\,\mathrm{d}a.
\label{eq:depletion}
\end{eqnarray}

Since we are often interested in the grain mass,
it will be convenient to consider the grain
size distribution per unit grain mass rather than
per unit grain radius. Thus, we define
$\tilde{n}(m,\, t)$ as the number density of grains
with mass between $m$ and $m+\mathrm{d}m$. The
two functions, $\tilde{n}$ and $n$, are related by
$\tilde{n}(m,\, t)\,\mathrm{d}m=n(a,\, t)\,
\mathrm{d}a$; that is, $\tilde{n}=n/(4\pi a^2s)$
by using equation (\ref{eq:mass}). Then, the time
evolution of $\tilde{n}$ is obtained from
equation (\ref{eq:continuity}) as
\begin{eqnarray}
\frac{\partial (m\tilde{n})}{\partial t}+\dot{m}
\frac{\partial (m\tilde{n})}{\partial m}=
\frac{1}{3}\dot{m}\tilde{n},\label{eq:dndt_acc1}
\end{eqnarray}
where we have used equation (\ref{eq:mass}),
$\dot{m}\equiv\mathrm{d}m/\mathrm{d}t=4\pi a^2s\dot{a}$,
$\partial /\partial m=[1/(4\pi a^2s)]\partial /\partial a$,
and $\partial\dot{a}/\partial a=0$.

It is convenient to define ${\sigma}$ as
\begin{eqnarray}
\sigma (m,\, t)\equiv m\tilde{n}(m,\, t).\label{eq:sigma}
\end{eqnarray}
Then, equation (\ref{eq:dndt_acc1}) is reduced to
\begin{eqnarray}
\frac{\partial\sigma}{\partial t}+\dot{\mu}
\frac{\partial\sigma}{\partial\mu}=\frac{1}{3}
\dot{\mu}\sigma,\label{eq:continuity_sigma}
\end{eqnarray}
where $\mu\equiv\ln m$ and
$\dot{\mu}\equiv\mathrm{d}\mu /\mathrm{d}t$.
Noting that $\dot{\mu}=3\dot{a}/a$, we obtain
from equation (\ref{eq:dadt})
\begin{eqnarray}
\dot{\mu}=\frac{3\xi (t)}{\tau (m)},\label{eq:dmudt}
\end{eqnarray}
where $\tau$ is now expressed as a function of
$m$ instead of $a$.
The evolution of $\xi$ is calculated by
(equations \ref{eq:tau}, \ref{eq:xi}, \ref{eq:depletion},
and \ref{eq:sigma})
\begin{eqnarray}
\frac{\mathrm{d}\xi}{\mathrm{d}t}=
\frac{-3f_\mathrm{X}\xi (t)}{m_\mathrm{X}n_\mathrm{X,tot}}
\int_0^\infty\frac{\sigma (m,\, t)}{\tau (m)}\,\mathrm{d}m.
\label{eq:dxidt}
\end{eqnarray}
We solve equations (\ref{eq:continuity_sigma}),
(\ref{eq:dmudt}), and (\ref{eq:dxidt}).
The discretized form of
equation (\ref{eq:continuity_sigma}) is shown in
Appendix \ref{app:continuity}. The initial
conditions are described in Section \ref{subsec:initial}.

\subsection{Coagulation}\label{subsec:coag}

For the evolution of grain size distribution by
coagulation, we apply the formulation developed
in our previous work \citep{hirashita09}.
Here we adopt an analytic formula derived by
\citet{ormel09} for the turbulent velocities. Below
we briefly overview the method, focusing on the
treatment of the grain velocities.

We consider thermal (Brownian) motion and
turbulent motion as a function of grain mass (or
radius, which is related to the mass by
equation \ref{eq:mass}). The turbulent motion
becomes important particularly for large grains
\citep{ossenkopf93,weidenschilling94,ormel09}.
The velocity as a function of grain mass $m$ is
given by a combination of thermal (Brownian) and
turbulent velocities ($v_\mathrm{th}$ and
$v_\mathrm{turb}$, respectively) as
\begin{eqnarray}
v(m)^2=v_\mathrm{th}(m)^2+v_\mathrm{turb}(m)^2.
\end{eqnarray}
{The thermal velocity is given by
$v_\mathrm{th}^2=8kT_\mathrm{gas}/(\pi m)$
\citep{spitzer78}, which is numerically evaluated
as}
\begin{eqnarray}
v_\mathrm{th} & = & 0.529\left(
\frac{T_\mathrm{gas}}{10~\mathrm{K}}\right)^{1/2}
\left(\frac{a}{0.1~\micron}\right)^{-3/2}\nonumber\\
& & \times \left(
\frac{s}{3~\mathrm{g~cm}^{-3}}
\right)^{-1/2}~\mathrm{cm~s}^{-1}.
\end{eqnarray}
The velocity driven by turbulence is given by
\begin{eqnarray}
v_\mathrm{turb} & = & 8.8\times 10^2\left(
\frac{T_\mathrm{gas}}{10~\mathrm{K}}\right)^{1/4}
\left(\frac{a}{0.1~\micron}\right)^{1/2}\nonumber\\
& \times & \left(
\frac{n_\mathrm{mol}}{10^5~\mathrm{cm}^{-3}}\right)^{-1/4}
\left(\frac{s}{3~\mathrm{g~cm}^{-3}}\right)^{1/2}~
\mathrm{cm~s}^{-1},\label{eq:v_turb}
\end{eqnarray}
where $n_\mathrm{mol}$ is the number density of
the molecular gas, which is related to
the number density of hydrogen nuclei,
$n_\mathrm{H}$, as
$n_\mathrm{mol}=n_\mathrm{H}/1.7$ for
the cosmic abundance
\citep{ormel09}\footnote{$n_\mathrm{mol}=1.7n_\mathrm{H}$
in \citet{ormel09} should be $n_\mathrm{mol}=n_\mathrm{H}/1.7$
(note that $n_\mathrm{mol}$ is denoted as $n$ in
\citealt{ormel09}).
{We can also easily check the validity of the
``intermediate regime'' by using the estimates
in \citet{ormel09}: with $T_\mathrm{gas}=10$ K
and $n_\mathrm{H}=10^3$ cm$^{-3}$,
$\mathrm{St}\simeq 1.3\times 10^{-2}(a/0.1~\micron)$
and $\mathrm{Re}^{-1/2}\simeq 4.6\times 10^{-4}$,
where St is the Stokes number and Re is the
Reynolds number. Thus, the condition for the intermediate
regime, $\mathrm{Re}^{-1/2}<\mathrm{St}<1$, is satisfied
in the size range where the turbulent motion is
dominant over the thermal motion.
\citet{yan04} adopted larger size and velocity for the
largest eddies, so Re is larger than assumed above.
In such a case, the intermediate regime is completely
valid for all grain sizes treated in this paper.}}.
In fact, equation (\ref{eq:v_turb}) represents
the relative velocity of equal-sized grains. We
adopt this as a typical velocity for the following reasons:
(i) a simple form for the velocity as a function of
grain size fits our approximate treatment of
the relative velocity (see below); (ii) the relative
velocity is determined by the velocity of the
larger grain, which is coupled with the larger-scale
turbulent motion, so the relative velocity
is always of the order of $v_\mathrm{turb}(a_1)$,
where $a_1$ is the radius of the larger grain;
(iii) the uncertainty caused by this rough treatment
of relative velocities,
compared with the analytic solution given by
equation (28) of \citet{ormel07}, is within a
factor of 1.3.

Grain velocities are dominated by thermal
motion for small grains ($a\la 0.002~\micron$),
while turbulence drives the motion of large grains
efficiently. If the grain size is too large, the grain
velocity becomes larger than the coagulation
threshold, which is calculated by the same way
as \citet{hirashita09}.
{Yet, the grain velocities ($\ll 1$ km s$^{-1}$)
are too small for shattering or erosion to occur
\citep{jones96}.}
For a grain colliding
with $a\sim 0.01~\micron$ grains, coagulation
occurs if the grain has a size smaller than
$\mbox{a few}\times 0.01~\micron$.
Each time-step is divided into four equal small
steps, and we
apply $v_{k\ell}=v_k+v_\ell$, $|v_k-v_\ell|$, $v_k$, and
$v_\ell$
{($v_k$ and $v_\ell$ are the grain
velocities evaluated at $a=a_k$ and $a_\ell$,
respectively, and
$v_{k\ell}$ is the relative velocity in the collision
between two grains in bins $k$ and $\ell$; see
Appendix \ref{app:coag} for
the discrete formulation)} in each step to consider a variety of
relative velocity directions \citep{jones94,hirashita09}.

\subsection{Initial conditions}\label{subsec:initial}

The initial grain size distribution is assumed to be
described by a power-law function with power
index $-r$ and
upper and lower bounds for the grain radii
$a_\mathrm{min}$ and $a_\mathrm{max}$, respectively:
\begin{eqnarray}
n(a,\, 0)=
{\displaystyle
\frac{(4-r)\rho_\mathrm{d}(0)}
{\frac{4}{3}\pi s(a_\mathrm{max}^{4-r}-a_\mathrm{min}^{4-r})}
}\, a^{-r}
\end{eqnarray}
for $a_\mathrm{min}\leq{a}\leq a_\mathrm{max}$.
If $a<a_\mathrm{min}$ or $a>a_\mathrm{max}$,
$n(a,\, 0)=0$. The dust mass density at $t=0$,
$\rho_\mathrm{d}(0)$ satisfies
equation (\ref{eq:rho_d}), and is given later by
equation (\ref{eq:rho_d0}). The dust mass density
is related to the initial condition for $\xi$ as
$\rho_\mathrm{d}(0)=(m_\mathrm{X}/f_\mathrm{X})
[1-\xi (0)]n_\mathrm{X,tot}$,
and the total number density of element X both in
gas and dust phases is written as
\begin{eqnarray}
n_\mathrm{X,tot}=\left(\frac{Z}{\mathrm{Z}_{\sun}}
\right)\left(\frac{\mathrm{X}}{\mathrm{H}}
\right)_{\sun}n_\mathrm{H},\label{eq:nxtot}
\end{eqnarray}
where $Z$ is the metallicity, and (X/H)$_{\sun}$ is
the solar abundance (the ratio of the number
of X nuclei to that of hydrogen nuclei at the solar
metallicity),
and $n_\mathrm{H}$ is the number density of
hydrogen nuclei. Therefore,
$\rho_\mathrm{d}(0)$ is related to the initial
condition for $\xi$ as
\begin{eqnarray}
\rho_\mathrm{d}(0)=
\frac{m_\mathrm{X}}{f_\mathrm{X}}[1-\xi (0)]
\left(\frac{Z}{\mathrm{Z}_{\sun}}\right)
\left(\frac{\mathrm{X}}{\mathrm{H}}\right)_{\sun}
n_\mathrm{H}.\label{eq:rho_d0}
\end{eqnarray}
We apply $Z=\mathrm{Z}_{\sun}$ (note that the
time-scales of both accretion and coagulation
are simply scaled as $Z^{-1}$). We assume
$\xi (0)=0.3$, which roughly matches the depletion
in the diffuse medium
\citep{savage96}. Although there is
uncertainty in the depletion
because of the assumed elemental abundance
pattern (usually the solar abundance pattern
is assumed), the following discussions on the
evolution of grain size distribution
are not altered as long as we adopt
a reasonable value such as $\xi (0)=0.1$--0.7.

\citet*{mathis77} show that the extinction curve in
the Milky Way can be fitted with a power-law grain
size distribution with $r=3.5$. Thus, we assume
that $r=3.5$. The effect of $r$ on accretion has
already been investigated by HK11.
Briefly, for larger $r$, small grains occupy
a larger fraction of total dust surface and the
total grain surface becomes larger; as a consequence,
the grain growth occurs more rapidly for larger $r$.
Since the largest grains are not susceptible to
accretion and coagulation as we show later, we fix
the maximum size as
$a_\mathrm{max}=0.25~\micron$ \citep{mathis77}.
The lower bound of the
grain size is poorly determined from the extinction
curve \citep{weingartner01}; thus,
we examine $a_\mathrm{min}=0.3$ and 1 nm.

\subsection{Selection of quantities}
\label{subsec:quantities}

As mentioned at the beginning of this section, we
consider silicate and graphite separately.
The quantities adopted in this paper are summarized
in Table \ref{tab:material}, and are based
on HK11. Since we are interested in
the interplay between accretion and coagulation,
we fix parameters which do not affect the relation
between those two processes.

\begin{table*}
\centering
\begin{minipage}{90mm}
\caption{Adopted quantities.}
\label{tab:material}
    \begin{tabular}{lccccc}
     \hline
     Species & X & $f_\mathrm{X}\,^\mathrm{a}$ &
     $m_\mathrm{X}$ [amu]\,$^\mathrm{b}$
     & (X/H)$_{\sun}\,^\mathrm{b}$ &
     $s$ [g cm$^{-3}$]\,$^\mathrm{c}$\\ 
     \hline
     Silicate & Si & 0.166 & 28.1 & $3.55\times 10^{-5}$ & 3.3 \\
     Graphite & C  & 1     & 12 & $3.63\times 10^{-4}$ & 2.26 \\
     \hline
    \end{tabular}
    
    \medskip

{    
$^\mathrm{a}$ For silicate, we assume a composition of
Mg$_{1.1}$Fe$_{0.9}$SiO$_4$ \citep{draine84}.\\
$^\mathrm{b}$ The atomic masses and the abundances
are taken from \citet{cox00}.\\
$^\mathrm{c}$ The material densities are taken from
\citet{draine84}.
}
\end{minipage}
\end{table*}

By using equation (\ref{eq:nxtot}), $\tau (a)$ in
equation (\ref{eq:tau})
can be estimated as
\begin{eqnarray}
\tau & = & 1.61\times 10^8\left(\frac{a}{0.1~\micron}
\right)\left(\frac{Z}{\mathrm{Z}_{\sun}}\right)^{-1}
\left(\frac{n_\mathrm{H}}{10^3~\mathrm{cm}^{-3}}
\right)^{-1}\nonumber\\
& & \times\left(\frac{T_\mathrm{gas}}{10~\mathrm{K}}
\right)^{-1/2}\left(\frac{S}{0.3}\right)^{-1}~
\mathrm{yr}
\label{eq:tau_sil}
\end{eqnarray}
for silicate, and
\begin{eqnarray}
\tau & = & 0.993\times 10^8\left(\frac{a}{0.1~\micron}
\right)\left(\frac{Z}{\mathrm{Z}_{\sun}}\right)^{-1}
\left(\frac{n_\mathrm{H}}{10^3~\mathrm{cm}^{-3}}
\right)^{-1}\nonumber\\
& & \times\left(\frac{T_\mathrm{gas}}{10~\mathrm{K}}
\right)^{-1/2}\left(\frac{S}{0.3}\right)^{-1}~
\mathrm{yr}
\label{eq:tau_gra}
\end{eqnarray}
for graphite. As mentioned in
Section \ref{subsec:initial}, we adopt
$Z=\mathrm{Z}_{\sun}$,
$n_\mathrm{H}=10^3$\,cm$^{-3}$ for the typical
values derived from observational properties
of Galactic molecular clouds \citep{hirashita00},
$T_\mathrm{gas}=10$ K
\citep{wilson97}, and $S=0.3$
\citep{leitch85,grassi11}. Both accretion and
coagulation have the same dependence of
time-scale on
metallicity and density as
$\propto (n_\mathrm{H}Z)^{-1}$

For $t$, we investigate a probable range for the
lifetime of molecular clouds. \citet{lada10}
mention that molecular clouds survive after
star formation activities lasting $\sim 2$~Myr.
The comparison with the age of stellar clusters
associated with molecular clouds indicates that
the lifetime of clouds is
$\sim 10$ Myr \citep{leisawitz89,fukui10}.
\citet{koda09} argue that molecular clouds
can be sustained over the circular time-scale
in a spiral galaxy ($\sim 100$ Myr). Therefore,
we examine $t\sim $ a few--100 Myr as a
probable range for the cloud lifetime.

\section{Results}\label{sec:result}

\subsection{Evolution of grain size distribution by
accretion and coagulation}\label{subsec:result_size}

\begin{figure*}
\includegraphics[width=0.45\textwidth]{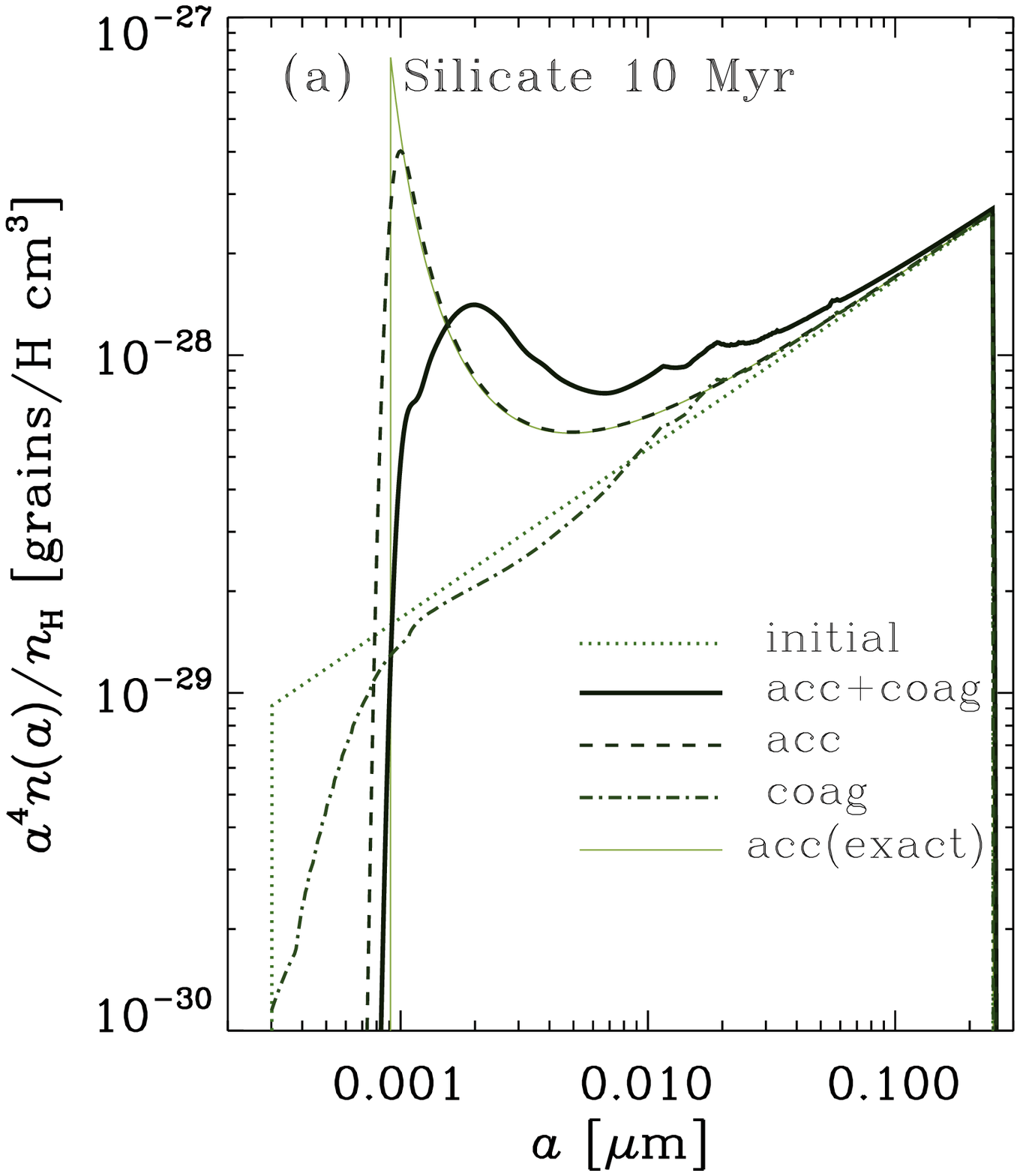}
\includegraphics[width=0.45\textwidth]{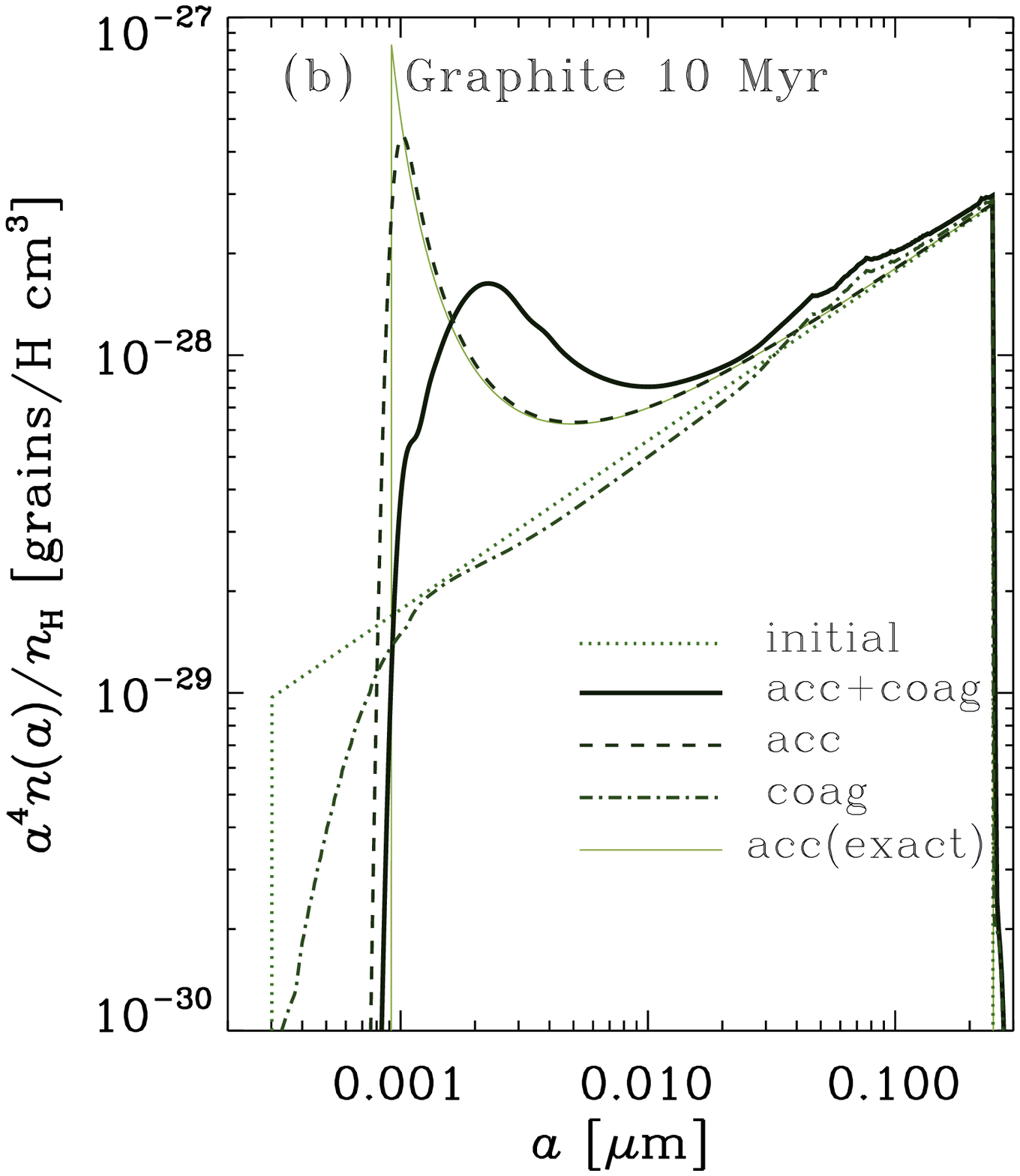}
 \caption{Grain size distributions in Model A at
 $t=10$ Myr. The thick solid, dashed, and dot-dashed
 lines show the results with both accretion and
 coagulation, only accretion, and only coagulation,
 respectively. The dotted line and the thin solid line
 present the initial size distribution and the analytic
 solution for accretion, respectively. Panels (a) and
 (b) show silicate and graphite, respectively.
 }
 \label{fig:acccoag}
\end{figure*}

We examine various cases, focusing on the interplay
between accretion and coagulation. We fix $r$ and
$a_\mathrm{max}$ since they do not affect the relative
role between accretion and coagulation
{(Section \ref{subsec:dependence})}.
Other parameters,
especially $a_\mathrm{min}$ and turbulent velocity, are
potentially important in this paper. For convenience,
we name each set of parameters Model A--D as shown in
Table \ref{tab:model}.

\begin{table}
\centering
\begin{minipage}{60mm}
\caption{Models.}
\label{tab:model}
    \begin{tabular}{ccccc}
     \hline
     Model & $r$ & $a_\mathrm{min}$ &
     $a_\mathrm{max}$ & turbulence\\
      & & [nm] & [$\micron$] &\\
     \hline 
     A & 3.5 & 0.3 & 0.25 & on\\
     B & 3.5 & 1 & 0.25 & on\\
     C & 3.5 & 0.3 & 0.25 & off\\
     D & 3.5 & 1 & 0.25 & off\\
     \hline
    \end{tabular}
\end{minipage}
\end{table}

In Fig.\ \ref{fig:acccoag}, we present the results
for Model A at $t=10$ Myr. To show the mass
distribution per logarithmic size, we multiply
${a}^4$ to ${n}$. Since the {growth}
rate of grain radius, $\dot{a}$, is independent
of $a$, the
impact of grain growth is significant at small
grain sizes. Moreover, gas-phase metals accrete
selectively onto small grains
because the grain surface
is dominated by small grains
\citep[see also][]{weingartner99}. The results
for silicate and graphite are similar
because the amount of available gas-phase
metals is similar.

In order to clarify the contributions from accretion
and coagulation, Fig.\ \ref{fig:acccoag} also shows
the cases where either only accretion or only
coagulation is taken into account in the
calculation. In Fig.\ \ref{fig:acccoag}, we observe
that both accretion and coagulation deplete
grains at $a\la 0.001~\micron$.
Accretion plays a significant role in increasing
the grain mass around $a\sim 0.001~\micron$,
and coagulation further pushes the peak
to $a\sim 0.002~\micron$ at 10 Myr.

Fig.\ \ref{fig:acccoag} also presents the analytical
solution for accretion calculated by the method in
HK11. Our numerical results with only accretion
match the analytical
solution very well except for the peak of the
size distribution around $a\sim 0.001~\micron$.
The sharp peak is smoothed because of the numerical
diffusion. Reproducing
the sharp peak is not important, however, since it
is smoothed out by coagulation in any case.

\begin{figure*}
\includegraphics[width=0.32\textwidth]{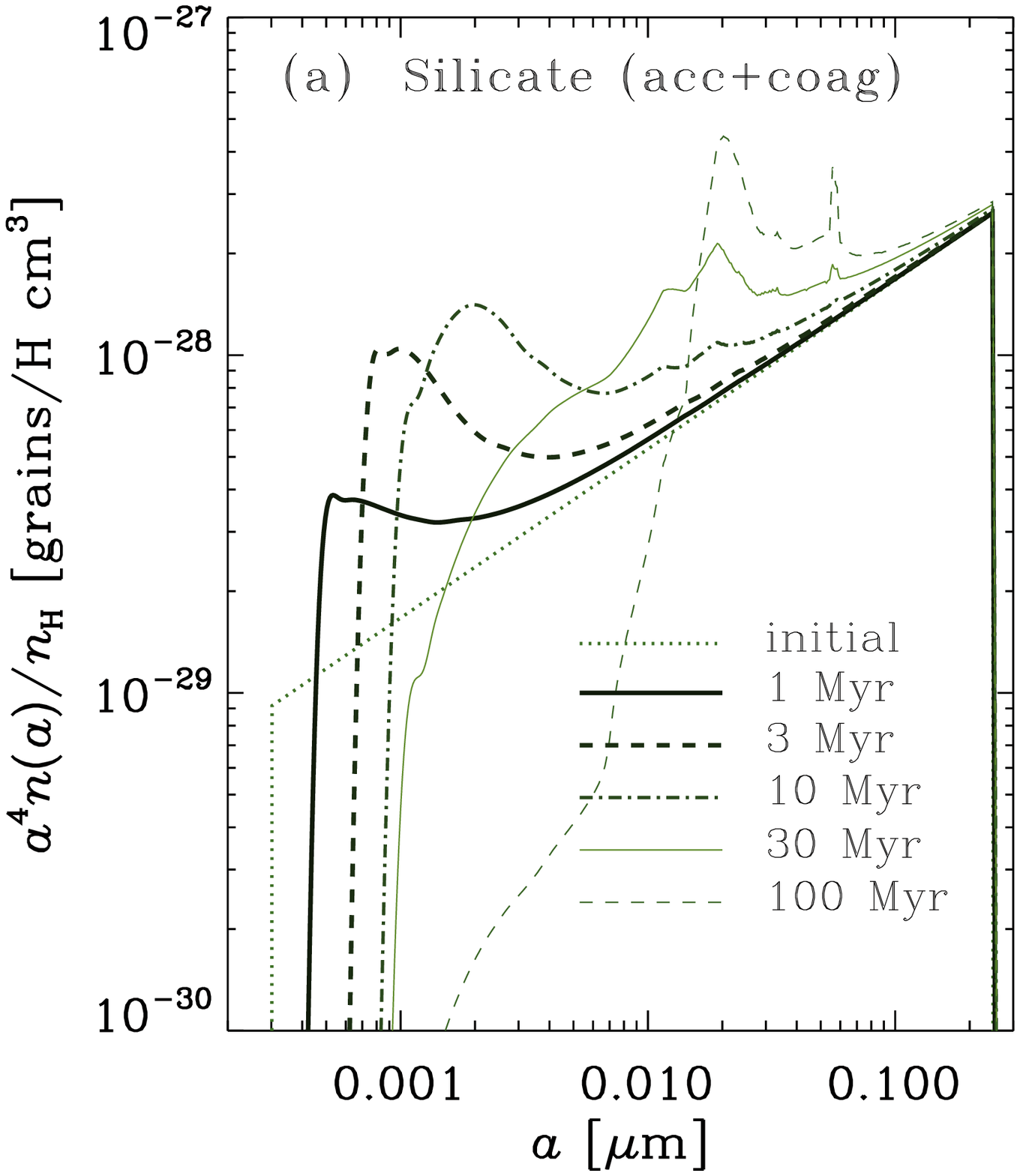}
\includegraphics[width=0.32\textwidth]{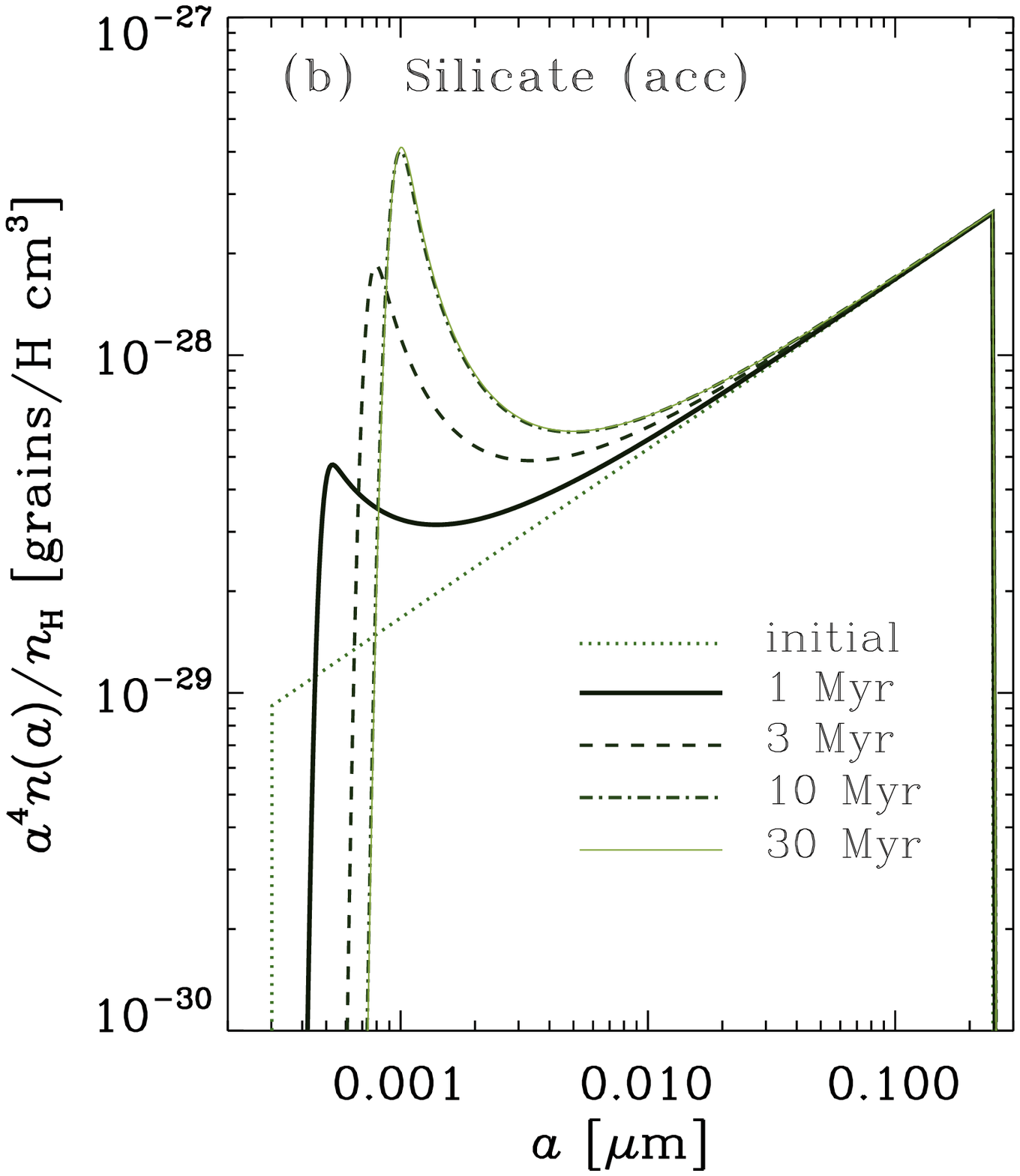}
\includegraphics[width=0.32\textwidth]{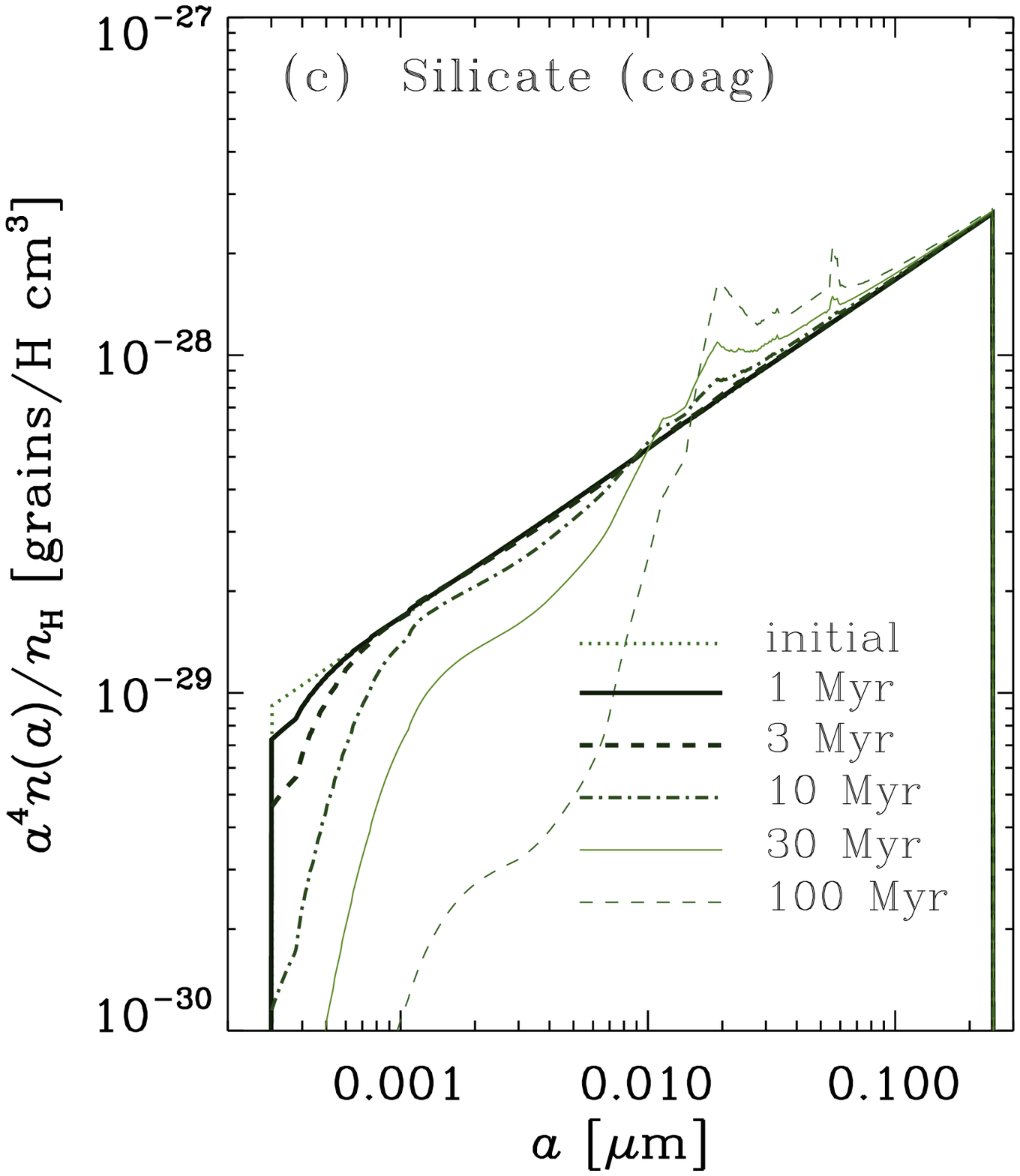}
 \caption{Same as Fig.\ \ref{fig:acccoag} but at various
$t$. The thick solid, thick dashed, dot-dashed,
thin solid, and thin dashed lines show the grain
size distributions at $t =1$, 3, 10, 30, and 100 Myr,
respectively (in Panel b, since the grain size
distributions are the same for $t\geq 10$ Myr, we
do not show the size distribution at $t=100$~Myr).
Each panel shows
the case with (a) both accretion and coagulation,
(b) only accretion, and (c) only coagulation.
 }
 \label{fig:acccoag_ev}
\end{figure*}

Next, we show the results for Model A at various $t$
in Fig.~\ref{fig:acccoag_ev}.
Fig.\ \ref{fig:acccoag_ev}a presents the grain
growth by accretion and coagulation. We observe
continuous grain growth. The spiky features around
$a\sim 0.01$--0.06 $\micron$
are produced by the accumulation
of grains at sizes where the grain velocity
reaches the coagulation threshold.
The discrete
spikes are artifact of our treatment of relative
velocities, which are evaluated in a discrete
way (Section \ref{subsec:coag}). If we consider
a smooth distribution of the
relative direction between the grains in collision,
those spiky features should
be smoothed out.

To analyze the results in
Fig.\ \ref{fig:acccoag_ev}a, we examine
the contribution from accretion and coagulation.
We show the results for only accretion and only
coagulation in Figs.~\ref{fig:acccoag_ev}b and c,
respectively. Comparing all the panels in
Fig.~\ref{fig:acccoag_ev}, we see that
the evolution of grain size distribution is first
driven by {both accretion and coagulation},
and is later caused by
coagulation. In fact, the accretion time-scale
is estimated by equation (\ref{eq:dadt}) as
$a/\dot{a}=\tau (a)/\xi (t)$, which is
$\sim 5$~Myr for $a\sim 0.001~\micron$
if we use the value of $\xi$ at $t=0$ ($\xi =0.3$).
Coagulation of small grains also occurs on a
similar time-scale.
{Indeed, according to equation (A5) in
\citet{hirashita_omukai}, the coagulation
time-scale can be estimated as
$t_\mathrm{coag}\sim 4
(n_\mathrm{H}/10^3~\mathrm{cm}^{-3})^{-1}
(a/10^{-3}~\micron )^{5/2}
(s/3~\mathrm{g~cm}^{-3})^{3/2}
(\mathcal{D}/3\times 10^{-3})^{-1}
(T/10~\mathrm{K})^{-1/2}$ Myr,
where $\mathcal{D}$ is the dust-to-gas ratio.
Although coagulation is as effective as accretion
in depleting the small-sized grains, it is
the role of accretion that creates the strong
bump around $a\sim 0.002~\micron$.}
As we observe in Fig.\ \ref{fig:acccoag_ev}b,
accretion stops at several Myr because of the
depletion of gas phase metals; thus, the evolution is
purely driven by coagulation after several Myr.
Comparing Figs.\ \ref{fig:acccoag_ev}a and c,
the typical grain size achieved by coagulation
is not sensitive to the presence of accretion,
but the bump around $\sim 0.02~\micron$
is stronger in the presence of accretion.
Thus, accretion helps to enhance the bump,
while the typical grain size
reached by coagulation is not affected by
accretion.

\begin{figure}
\includegraphics[width=0.45\textwidth]{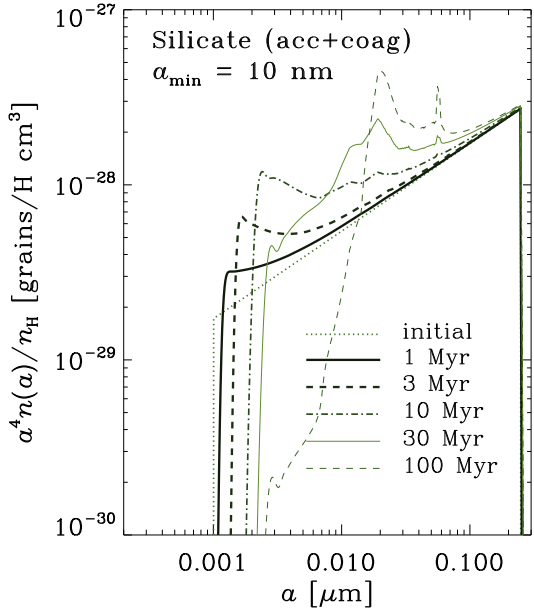}
 \caption{Same as Fig.\ \ref{fig:acccoag_ev}a but for a larger
 minimum grain radius, $a_\mathrm{min}=1$~nm (Model B).
 The thick solid, thick dashed, dot-dashed, thin solid, and
thin dashed lines show the cases for $t =1$, 3, 10, 30, and
100 Myr, respectively.
 }
 \label{fig:acccoag_ev_amin}
\end{figure}

Since grain growth predominantly occurs at
small sizes, it is expected that grain growth
is sensitive to $a_\mathrm{min}$. In
Fig.\ \ref{fig:acccoag_ev_amin}, we show the
evolution of grain size distribution for
Model B, in which we assume
$a_\mathrm{min}=10$ nm instead of 3 nm.
In this model, grain growth by accretion lasts
3 times longer than in Model A, because of
3 times larger $a_\mathrm{min}$ (note that the
grain growth time-scale is proportional to $a$;
equation \ref{eq:tau}). However, the
growth of large grains (typically
$a\ga 0.003~\micron$)
is not affected by the change of $a_\mathrm{min}$.

Regardless of $a_\mathrm{min}$, turbulence has
a significant impact on the grain size distribution
at $a\ga 0.003~\micron$ through coagulation,
especially after accretion is terminated by the
depletion of gas-phase metals. In this paper, we
have followed \citet{ormel09} for the grain motion
driven by turbulence. \citet*{yan04} show that the
grain motion in magnetized turbulence can also
be excited by gyroresonance. However, the grain
motion in dense medium is dominated by
hydrodynamical drag even in their models, and
the velocities that they calculated are similar to
ours for $a\la$ a few $\times 10^{-2}~\micron$.

\begin{figure}
\includegraphics[width=0.45\textwidth]{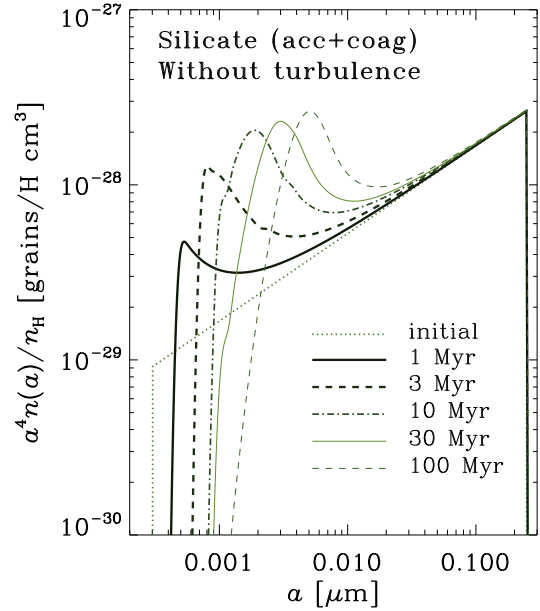}
 \caption{Same as Fig.\ \ref{fig:acccoag_ev}a but we show
 the case of no turbulence (Model~C).
 }
 \label{fig:acccoag_noturb}
\end{figure}

In order to examine the significance of turbulent
motion, we also calculate the case where the
turbulent velocity component is neglected; that is,
$v=v_\mathrm{th}$ (Model C). We show
the result in Fig.~\ref{fig:acccoag_noturb}.
Comparing Fig.\ \ref{fig:acccoag_noturb} with
Fig.\ \ref{fig:acccoag_ev}a, we observe that
the effect of turbulence on the grain size distribution
appears after $t\sim 10$ Myr, when coagulation
begins to be important for $a\ga 0.006~\micron$;
for this grain size range, the grain motion is
predominantly driven by turbulence. For
$t<10~\mathrm{Myr}$, however, turbulence does
not significantly affect the grain size distribution.

\subsection{Effect of coagulation on the grain mass
growth}\label{sec:mass_growth}

Grain growth by accretion in interstellar clouds is
one of the most important processes
that govern the grain mass budget in an entire
galactic system. Following HK11, we
define the increased fraction of dust mass in
a cloud, $\beta$, as
\begin{eqnarray}
\beta =
\frac{\rho_\mathrm{d}(\tau_\mathrm{cl})}
{\rho_\mathrm{d}(0)}-1,\label{eq:beta}
\end{eqnarray}
where $\tau_\mathrm{cl}$ is the lifetime of
clouds hosting the grain
growth.\footnote{In HK11,
only accretion is considered, so
$\rho_\mathrm{d}$ is proportional to the
mean value of $a^3$. However, if we take
coagulation into account, the grain number
density also changes. Thus, in this paper,
we cannot write $\beta$ by using the mean
value of $a^3$.}
The dust mass in the cloud becomes $(\beta +1)$
times as much as the initial
value at the cloud lifetime, when the dust
grown in the cloud returns in the diffuse ISM.
By using $\beta$, the
increasing rate of dust mass in a galaxy
should be written as
\begin{eqnarray}
\left[\frac{\mathrm{d}M_\mathrm{dust}}{\mathrm{d}t}
\right]_\mathrm{acc}=
\frac{\beta X_\mathrm{cl}M_\mathrm{dust}}
{\tau_\mathrm{cl}},\label{eq:dMdt2}
\end{eqnarray}
where $M_\mathrm{dust}$ is the total dust mass 
in the galaxy, and $X_\mathrm{cl}$ is the mass
ratio of clouds hosting the grain growth to the
total gas mass (HK11).

\begin{figure}
\includegraphics[width=0.45\textwidth]{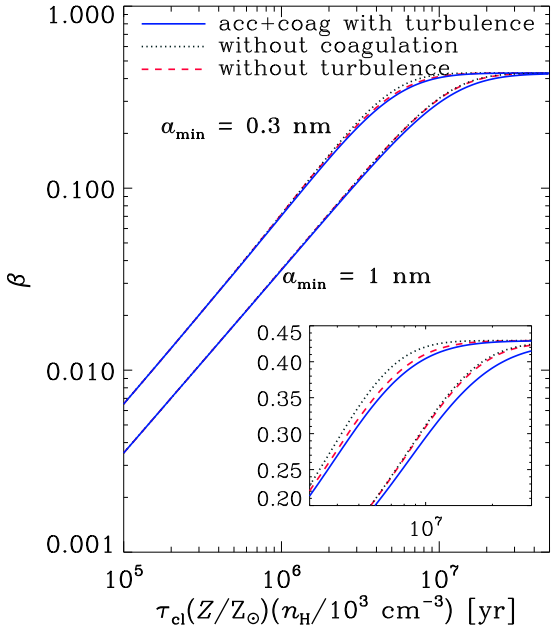}
 \caption{Increment of dust mass, $\beta$, defined
in equation (\ref{eq:beta}) as a function of the
cloud lifetime $\tau_\mathrm{cl}$. To clarify the
dependence on metallicity and density, the scaling
with $Z$ and $n_\mathrm{H}$ is also
shown in the horizontal axis.
{The difference in the upper and lower lines
with the same line species is the value of
$a_\mathrm{min}$ (0.3 nm for the upper lines
and 1 nm for the lower lines).
The solid, dotted,
and dashed lines show the results with
accretion and coagulation in the presence of
turbulence, without coagulation (i.e.\ only
accretion), and without turbulence (but with
both accretion and coagulation considered),
respectively. Note that the upper solid line,
the upper dashed line, the lower solid line,
and the lower dashed lines are equivalent to
Models A, C, B, and D, respectively.}
In the small window, we zoom in to clarify the
distinction between the lines (vertical axis
is on linear scale instead of logarithmic scale).
 }
 \label{fig:rhod}
\end{figure}

Noting that the time-scales of accretion and
coagulation are both scaled with the density of
metals [$\propto (n_\mathrm{H}Z)^{-1}$], $\beta$
can be regarded as a function of
$\tau_\mathrm{cl}n_\mathrm{H}Z$.
In Fig.\ \ref{fig:rhod}, we show $\beta$ for
Models A and B to examine the dependence
on grain size distribution
($a_\mathrm{min}=0.3$ nm and 1 nm,
respectively). Because accretion occurs more
efficiently at smaller grain sizes, $\beta$ is larger
for $a_\mathrm{min}=0.3$ nm than for
$a_\mathrm{min}=1$ nm. Accretion saturates if
the metals in gas phase
are used up. Thus,
$\beta\to \xi (0)/[1-\xi (0)]$
if the cloud lifetime is long enough.

In Fig.\ \ref{fig:rhod}, we also plot the cases
where we neglect coagulation (dotted
lines).\footnote{We have confirmed that our results
without coagulation matches
the analytical results by HK11.}
There
is a slight indication that coagulation suppresses
accretion. This is explained as follows: Coagulation
pushes grains to larger sizes
with the total volume conserved so that the
surface-to-volume ratio of the grains decreases.
Note that accretion rate is predominantly regulated
by the surface-to-volume ratio.
Nevertheless this suppression effect is very
small and we can conclude that
coagulation has little impact on accretion. This
also means that we can neglect coagulation as long
as we are interested in the total dust mass in
a galaxy, although we should keep in mind that
coagulation can affect the grain size distribution
(Section \ref{subsec:result_size}).

Finally, we also show the case where turbulence is
neglected (Models C and D) in Fig.\ \ref{fig:rhod}.
We observe that the effect of turbulence on
$\beta$ is negligibly small. If $a_\mathrm{min}$
is larger, coagulation is very inefficient if
there is no turbulent motion; thus, the
grain growth without turbulence is practically the
same as the grain growth without coagulation. We
can also conclude that turbulence has little impact
on the grain mass increase in clouds, although we
should note that turbulence can affect the grain
size distribution (Section \ref{subsec:result_size}).

\subsection{Dependence on other parameters}
\label{subsec:dependence}

We briefly discuss the influence of various
parameters. The dependence of the accretion
time-scale on physical quantities are clarified
in equations (\ref{eq:tau_sil})
and (\ref{eq:tau_gra}). Both accretion and
coagulation follow scaling of
time-scale $\propto (Zn_\mathrm{H})^{-1}$,
so that we get
the same grain size distribution at the
same $tZn_\mathrm{H}$. Precisely, the
grain velocities driven by turbulence also
depends on $n_\mathrm{H}$, but this
dependence is
only $\propto n_\mathrm{H}^{1/4}$.
The time-scale of grain growth is also
inversely proportional to the sticking efficiency.
Regarding
the gas temperature $T_\mathrm{gas}$,
both accretion
and coagulation are again regulated by
the same scaling $\propto T_\mathrm{gas}^{-1/2}$
if coagulation is purely driven by thermal
motion. Nevertheless, the gas temperature in cold
gas is at most $\sim 100$ K \citep{wilson97};
thus, the temperature just causes a factor of
3 difference in the time-scales. Coagulation
driven by turbulence is little affected by
$T_\mathrm{gas}$, because the dependence
of the turbulence-driven grain velocity on
gas temperature is weak
($\propto T_\mathrm{gas}^{1/4}$).

As mentioned in Section \ref{subsec:initial},
we assume that $r=3.5$. If $r$ is smaller/larger,
the number of grains at the smallest sizes is
smaller/larger. Thus, as already shown in HK11
(see their Fig.\ 3), accretion occurs more
slowly/quickly. Because coagulation also occurs
at the smallest grain sizes, the relative role
of accretion and coagulation does not change
even if we change $r$. We should note that
$r=3.5$ is supported from the extinction curves in
the Milky Way and Magellanic Clouds
\citep{mathis77,pei92}. Some theoretical
studies also indicate that disruption or shattering
of grains in the diffuse ISM processes the
grain size distribution to
a power-law with $r\simeq 3.5$
\citep[e.g.][]{hellyer70}.

As shown above, grains at
$a\sim a_\mathrm{max}$ are not affected by
accretion and coagulation. Thus, changing
$a_\mathrm{max}$ does not affect our results.
This is because both accretion and coagulation
occurs efficiently for grains which have a large
surface-to-volume ratio.

\section{Discussion}\label{sec:discussion}

Now we discuss two issues related to accretion
and coagulation. One is effects of these
processes on the extinction curve, and the other
is an implication for galaxy evolution.

\subsection{Effects on the extinction curve}
\label{subsec:ext}

Following \citet{hirashita09} and
\citet{odonnell97}, we examine how the grain size
evolution affects the extinction curve.
\citet{hirashita09} adopt an initial condition
which reproduces the Milky Way extinction curve,
and incorporate the
change of the grain size distribution by
shattering and coagulation to examine how the
extinction curve is modified
by these processes. Here we examine effects of
accretion and coagulation on the extinction curve.

The initial condition is set up so that the grain
abundance is consistent with the mean
extinction curve of the Milky Way by \citet{pei92};
that is, we adopt the same abundances of C and
Si as listed in Table \ref{tab:material} but apply
$\xi (0)=0.25$ and 0.15 for silicate and graphite,
respectively, under Model A. This rough agreement
is sufficient for our aim, since our aim is not detailed
fitting of the extinction curve. We adopt the
same dust optical properties and calculation method
of extinction curves as in \citet{hirashita09}:
The grain extinction cross section as a function
of grain size is derived
from the Mie theory, and is weighted with the grain
size distribution to obtain an extinction curve.

\begin{figure}
\includegraphics[width=0.45\textwidth]{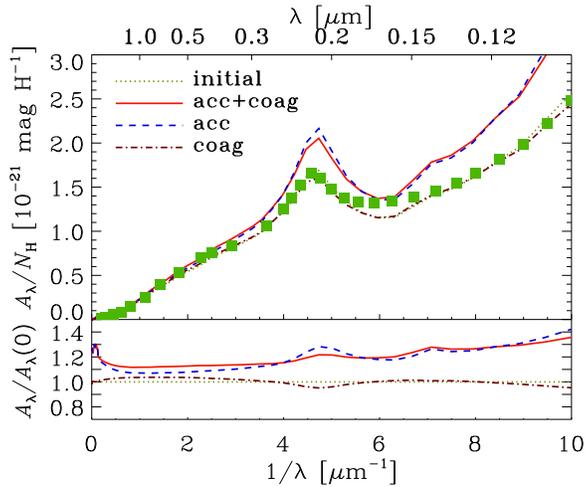}
 \caption{Upper panel: Extinction curves in units of
magnitude per hydrogen. The dotted line (almost
identical to the dot-dashed line) is the initial
extinction curve before accretion and coagulation.
The solid, dashed, and dot-dashed lines show the
results with both accretion and coagulation, only
accretion, and only coagulation, respectively, at
$t=10$ Myr, based on the parameter sets in
Model A. The filled squares represent the
Galactic extinction data taken from \citet{pei92}.
Lower Panel: The extinction divided by the initial
extinction. The correspondence between the
line types and the models are the same as above.
 }
 \label{fig:extinc}
\end{figure}

In Fig.\ \ref{fig:extinc}, we show the calculated
extinction curves in units of magnitude per
hydrogen. We present the results with both
accretion and coagulation,
only accretion, and only coagulation at $t=10$ Myr.
First, we observe that accretion increases the
grain opacity at all wavelengths because the grain
mass grows. In order to examine the wavelength
dependence, we also show the ratio between the
extinction curve at $t=10$ Myr to that at
$t=0$ Myr (initial) in the lower panel.
We find that accretion  `steepens' the extinction
curve rather than `flattens' it. Naively
it may be expected that
the extinction curve would be flattened after
accretion, because
the mean grain size becomes large. Contrary
to this expectation, the extinction curve becomes
steeper after accretion.
This is explained as follows. The extinction at a
wavelength $\lambda$ is proportional to
$a^3$ if $a\ll\lambda$, while it is proportional to
to $a^2$ (i.e., less sensitive to the grain size) if
$a\ga\lambda /(2\pi )$ \citep{bohren83}.
In other words, the ultraviolet (UV) extinction is
more sensitive to the enhancement of small grains
than the extinction at longer wavelengths.
Therefore, the extinction at shorter wavelengths
increases more sensitively as a result of accretion.
The 0.22-$\micron$ bump, which is due to small
graphite, also becomes more prominent
after accretion. At optical and
near-infrared wavelengths, the extinction is
relatively insensitive compared to other wavelengths,
since it is more affected by the largest grains intact
after accretion and coagulation.
As $\lambda$ becomes large at mid-infrared
and longer wavelengths
(i.e.\ $\lambda\gg 2\pi a_\mathrm{max}$), the
extinction is just
proportional to the grain mass (or volume).

As shown in Fig.\ \ref{fig:extinc} (comparison
between the solid and dashed lines), coagulation
flattens the UV extinction curve and lowers
the carbon bump. This is because of the depletion
of small grains. Coagulation conserves
the total grain mass (or volume); thus, the
extinction curve
at $\lambda\gg 2\pi a_\mathrm{max}$ is unaffected
by coagulation.

Observations of the Milky Way show that extinction
curves are flatter in the directions of dense
clouds \citep{mathis81,cardelli89}. This is not
consistent with accretion, but is consistent with
coagulation. \citet{cardelli89} also show that the
extinction per hydrogen nucleus
is smaller toward dense clouds than in the diffuse
ISM. This requires strong coagulation. The cloud
lifetime is possibly larger than 10 Myr,
so that coagulation can push the
grains to large sizes. Indeed, \citet{koda09}
suggest a lifetime of $\sim 100$ Myr or longer.
Another possibility of strong coagulation is
that coagulation additionally occurs in extremely
dense regions,
e.g.\ regions associated with star formation
{\citep{ormel09,hirashita_omukai}}.

\subsection{Significance in galaxy evolution}

Grain growth by accretion is shown to be
important not only in nearby galaxies, but also
in some high-redshift galaxy populations.
As shown in previous studies
(HK11; \citealt{inoue12,asano12}),
grain growth by accretion becomes prominent
if the metallicity exceeds a critical value. Those
studies pointed out general importance of
grain growth by accretion in nearby galaxies.
It is also indicated that grain growth is indeed
governing the dust abundance
in distant quasars
\citep{michalowski10,pipino11,valiante11,asano12}
\citep[but see][]{gall11b}.
In these objects, it is expected that
coagulation is also occurring, but according to
our results in this paper, the effect of
coagulation on {the total dust content} can be
neglected. If we are
interested in the evolution of grain
size distribution, however, we should take
coagulation into account, especially
after a significant fraction of gas-phase
metals are locked into dust grains.

Finally, in Table \ref{tab:ext_illust}, we briefly
overview the effects of various
interstellar processes on the extinction curve,
including those
treated in our previous papers. The carbon bump
and the UV slope are enhanced by shattering
\citep{hirashita09} and accretion (this paper),
while they are flattened by coagulation
(\citealt{hirashita09}; this paper) and
shock destruction \citep{nozawa07,hirashita08}.
We should
consider at least these effects as processes
determining the shape of the extinction curves
observed in a variety of galaxies including
high-redshift quasars and gamma-ray bursts
\citep[e.g.][for recent observations]{gallerani10,jang11}.

\begin{table}
\centering
\begin{minipage}{80mm}
\caption{Summary of effects on the extinction curve.}
\label{tab:ext_illust}
    \begin{tabular}{lccc}
     \hline
     Process & UV slope$^\mathrm{a}$ &
     Carbon bump$^\mathrm{b}$ & ref.$^\mathrm{c}$\\ 
     \hline 
     Accretion & $+$ & $+$ & 1\\
     Coagulation & $-$ & $-$ & 1, 2\\
     Shattering & $+$ & $+$ & 2\\
     Shock destruction & $-$ & $-$ & 3\\
     \hline
    \end{tabular}

\medskip

$^\mathrm{a}$ $+$: steeper; $-$: flatter.\\
$^\mathrm{b}$ $+$: stronger; $-$: weaker.\\
$^\mathrm{c}$ 1) this paper; 2) \citet{hirashita09};
3) \citet{hirashita08}.
\end{minipage}
\end{table}

\section{Conclusion}\label{sec:conclusion}

In this paper, we have focused on grain growth
processes in interstellar clouds. We have
formulated and calculated the evolution
of grain size distribution by accretion and
coagulation in an interstellar cloud.
Destructive processes are neglected in this
paper, since they are considered to be ineffective
in dense regions.
We have confirmed the previous analytic
results that grains smaller than
$\sim 0.001~\micron$ are depleted by accretion on
a time-scale of several Myr. Coagulation also occurs
on a similar time-scale,
but accretion has a
prominent impact in making a large bump in the
grain size distribution
around $a\sim 0.001~\micron$ because of
the grain mass increase.
After a major part of the gas phase metals
are used up, accretion stops and
coagulation is the only mechanism that
pushes the grains toward larger sizes.
It is only at this stage that grain
motion driven by
turbulence can affect the evolution of grain size
distribution. Regardless of whether accretion
takes place or not, the typical grain size
achieved by coagulation is similar.

We have also examined the grain mass increase
during the cloud lifetime by introducing $\beta$
in equation (\ref{eq:beta}) such that the
grain mass becomes $(1+\beta )$ times as much as
the initial value. 
We have found that coagulation
slightly suppresses accretion although
this effect is not significant and can be neglected
in discussing the evolution of the total grain mass
in galaxies. In other words,
coagulation can be neglected
in discussing the evolution of the total grain mass
in galaxies.

Finally, we have investigated the effects of accretion
and coagulation on the extinction curve. We have
found that accretion enhances the UV slope and
the carbon bump in spite of the increases in the mean
grain radius. As expected, coagulation flattens these
features.

\section*{Acknowledgments}
We are grateful to an anonymous referee for helpful
comments, which greatly improved the discussion
and content of this paper.
We thank T. Nozawa, N. Scoville, J. Koda, and P. Capak
for helpful discussions on dust and molecular clouds.
H.H. is supported by NSC grant 99-2112-M-001-006-MY3.

\appendix

\section{Discrete formulation}\label{app:discrete}

For numerical calculation, we consider $N$ discrete
grain radii, and denote the lower and upper
bounds of the $i$th ($i=1,\,\cdots ,\, N$) bin as
$a_{i-1}^\mathrm{(b)}$
and $a_i^\mathrm{(b)}$, respectively. We adopt
$a_{i}^\mathrm{(b)}=a_{i-1}^\mathrm{(b)}\delta$,
$a_0^\mathrm{(b)}=a_\mathrm{l}$, and
$a_N^\mathrm{(b)}=a_\mathrm{u}$ with
$\log\delta =(1/N)\log (a_\mathrm{u}/a_\mathrm{l})$.
We represent the grain radius and mass in the $i$th bin
with
$a_i\equiv (a_{i-1}^\mathrm{(b)}+a_i^\mathrm{(b)})/2$
and $m_i\equiv (4\pi /3)a_i^3s$. The
boundary of the mass bin is defined as
$m_i^\mathrm{(b)}\equiv (4\pi /3)[a_i^\mathrm{(b)}]^3s$.
The interval of logarithmic mass grids is denoted as
$\Delta\mu =3\delta$. We also discretize the time as
$t_n=n\Delta t$. We use integer indexes $i$ and $n$ to
specify the discrete grain size and time, respectively.
We adopt $N=512$,
$a_\mathrm{l}=2\times 10^{-4}~\micron$,
and $a_\mathrm{u}=0.3~\micron$. We apply
$n(a_0,\, t)=n(a_N,\, t)=0$ for the boundary
condition.

\subsection{Accretion}
\label{app:continuity}

Now we explain how to discretize
equation (\ref{eq:continuity_sigma}). We denote the
value of quantity $Q$ at a discrete grid as $Q_i^n$
(recall that $i$ and $n$ specify grain-size
and temporal grids, respectively). Then, we
obtain the following equation as a discrete version
of equation (\ref{eq:continuity_sigma}):
\begin{eqnarray}
\frac{\sigma_i^{n+1}-\sigma_i^n}{\Delta t}+
\dot{\mu}_i^n\frac{\sigma_i^n-\sigma_{i-1}^n}{\Delta\mu}
=\frac{1}{3}\dot{\mu}_i^n\sigma_i^n,\label{eq:discrete}
\end{eqnarray}
where the difference is evaluated based on
upwind differencing. In hydrodynamical simulations,
more elaborate differencing methods have been developed,
but the simple difference is enough for the grain size
distribution (for example, we are not interested in
`shocks' or `instabilities' when we are treating the
grain size distribution in this paper).
We apply Courant condition
$\Delta t<\Delta\mu /\max_i(\dot{\mu}_i^n)$
for every time step. From equation (\ref{eq:dmudt}),
we observe that the minimum value of $\mu$ is
realized at the maximum value of $\tau (m)$, which
is small when $m$ is small (equation \ref{eq:tau}).
Therefore, the above Courant condition is
practically evaluated at $i=1$.

Solving equation (\ref{eq:discrete}) for
$\sigma_i^{n+1}$, we obtain
\begin{eqnarray}
\sigma_i^{n+1}=\sigma_i^n\left( 1-\dot{\mu}_i^n
\frac{\Delta t}{\Delta\mu}+\frac{\Delta t}{3}
\dot{\mu}_i^n \right) +\dot{\mu}_i^n
\frac{\Delta t}{\Delta\mu}\sigma_{i-1}^n.
\end{eqnarray}

\subsection{Coagulation}\label{app:coag}

The mass density of grains contained in the $i$th
bin, $\tilde{\rho}_i^n$ ($n$ is the time step), is
defined as
$\tilde{\rho}_i^n\equiv\sigma_i^nm_i\Delta\mu$.
The time evolution of $\tilde{\rho}_i$ by
coagulation can be written as
\begin{eqnarray}
\frac{\tilde{\rho}_i^{n+1}-\tilde{\rho}_i^n}{\Delta t}
& \hspace{-3mm}= & \hspace{-3mm}
-m_i\tilde{\rho}_i\sum_{\ell =1}^{N}\alpha_{\ell i}
\tilde{\rho}_\ell+\sum_{j=1}^{N}
\sum_{\ell =1}^N\alpha_{\ell j}
\tilde{\rho}_\ell
\tilde{\rho}_jm_\mathrm{coag}^{\ell j}(i),\nonumber\\
\end{eqnarray}
and
\begin{eqnarray}
\alpha_{\ell k} = \left\{
\begin{array}{ll}
{\displaystyle
\frac{\beta\sigma_{\ell k}v_{\ell k}}{m_km_\ell}
} &
\mbox{if $v_{\ell k}<v_\mathrm{coag}^{\ell k}$,} \\
0 & \mbox{otherwise,}
\end{array}
\right.
\end{eqnarray}
where $\beta$ is the sticking probability
{for coagulation}, and the
coagulated mass $m_\mathrm{coag}^{\ell j}(i)$
is determined as follows:
$m_\mathrm{coag}^{\ell j}(i)=m_\ell$ if
$m_{i-1}^\mathrm{(b)}\leq m_\ell +m_j<
m_i^\mathrm{(b)}$;\footnote{There was a typo in
\citet{hirashita09}.}
otherwise $m_\mathrm{coag}^{\ell j}(i)=0$. Coagulation
is assumed to occur only if the relative velocity is less
than the coagulation
threshold velocity $v_\mathrm{coag}^{\ell k}$
based on \citet{chokshi93}, \citet{dominik97},
and \citet{yan04}
\citep[see][for further details]{hirashita09}.
The cross-section for the coagulation is
$\sigma_{\ell k}=\pi (a_\ell +a_k)^2$.
We assume $\beta =1$ for the sticking probability.

{
To confirm the validity of our numerical scheme
of coagulation, we checked in \citet{hirashita09}
that the total dust mass is conserved.
We  also observe in \citet{hirashita_omukai}
that coagulation proceeds on an analytically
estimated time-scale: this check is roughly
equivalent with that in
the appendix D of \citet{brauer08}.
}

\bsp

\label{lastpage}

\end{document}